\documentclass{Interspeech}

\interspeechcameraready

\title{Listen through the Sound:\\ Generative Speech Restoration Leveraging Acoustic Context Representation}

\author[affiliation={}]{Soo-Whan}{Chung}
\author[affiliation={}]{Min-Seok}{Choi}

\affiliation[nocounter]{}{NAVER Cloud}{South Korea}
\email{soowhan.chung@navercorp.com}

\usepackage{comment}

\usepackage{overpic}
\usepackage{enumitem} 
\usepackage{overpic} 
\usepackage{color}
\usepackage[dvipsnames]{xcolor}
\usepackage{multirow,multicol}

\usepackage{pifont}
\usepackage{amsfonts}
\usepackage{booktabs}
\usepackage{tabularx}

\usepackage{graphicx}
\usepackage{subfig}
\usepackage{pgfplots, mathtools}

\usepackage{caption}
\definecolor{turquoise}{cmyk}{0.65,0,0.1,0.3}
\definecolor{purple}{rgb}{0.65,0,0.65}
\definecolor{dark_green}{rgb}{0, 0.5, 0}
\definecolor{orange}{rgb}{0.8, 0.6, 0.2}
\definecolor{red}{rgb}{0.8, 0.2, 0.2}
\definecolor{darkred}{rgb}{0.6, 0.1, 0.05}
\definecolor{blueish}{rgb}{0.0, 0.3, .6}
\definecolor{light_gray}{rgb}{0.7, 0.7, .7}
\definecolor{pink}{rgb}{1, 0, 1}
\definecolor{greyblue}{rgb}{0.25, 0.25, 1}

\newcommand{\PreserveBackslash}[1]{\let\temp=\\#1\let\\=\temp}
\newcolumntype{C}[1]{>{\PreserveBackslash\centering}p{#1}}
\newcolumntype{R}[1]{>{\PreserveBackslash\raggedleft}p{#1}}
\newcolumntype{L}[1]{>{\PreserveBackslash\raggedright}p{#1}}

\newcommand{\Fig}[1]{Fig.~\ref{fig:#1}}

\newcommand{\Tab}[1]{Tab.~\ref{tab:#1}}

\usepackage{blindtext}

\renewcommand{\paragraph}[1]{\vspace{1pt}\noindent\textbf{#1}\textbf{.}}
\newcommand{\scriptnote}[1]{\footnote{\scriptsize{#1}}}

\newcommand{\eg}{\textit{e.g.~}}
\newcommand{\ie}{\textit{i.e.~}}

\newcommand{\shortname}{ACX}

\newcommand{\furl}[1]{\scriptnote{\url{#1}}}

\newcommand{\xubsection}[1]{\vspace{-2pt}\subsection{#1}\vspace{-3pt}}

\newcommand{\xcaption}[1]{\vspace{0pt}\caption{#1}\vspace{-10pt}}
\newcommand{\tcaption}[1]{\vspace{0pt}\caption{#1}\vspace{-15pt}}

\begin{document}

\maketitle

\begin{abstract}\vspace{-4pt}
This paper introduces a novel approach to speech restoration by integrating a context-related conditioning strategy. Specifically, we employ the diffusion-based generative restoration model, UNIVERSE++, as a backbone to evaluate the effectiveness of contextual representations. We incorporate acoustic context embeddings extracted from the CLAP model, which capture the environmental attributes of input audio. Additionally, we propose an Acoustic Context (ACX) representation that refines CLAP embeddings to better handle various distortion factors and their intensity in speech signals. Unlike content-based approaches that rely on linguistic and speaker attributes, ACX provides contextual information that enables the restoration model to distinguish and mitigate distortions better. Experimental results indicate that context-aware conditioning improves both restoration performance and its stability across diverse distortion conditions, reducing variability compared to content-based methods.

\end{abstract}

\vspace{-4pt}
\section{Introduction}\vspace{-2pt}

In our daily lives, when we record speech, we encounter not only the speech itself but also unwanted background noise and reverberation from the environment.
At times, we face additional degradations, such as spectral distortion, narrow bandwidths, and even undesired amplitude clipping, depending on the recording environment or the device used.
In the past, to obtain high-quality speech from input signals, each distortion was addressed with a separate model, then handled in a cascading manner~\cite{zhao2018two,strake2019separated}.
However, recent advances in complexity and methodology of deep learning have led to integrated approaches that process multiple distortions together.
Early discriminative methods~\cite{hsieh2020wavecrn,zhao2020noisy} focused on estimating clean speech features (\eg spectrogram, mel-spectrogram), much like traditional speech enhancement methods.
When generative adversarial networks~(GANs) emerged, they~\cite{liu2022voicefixer,kim2023hd} further boosted the performance of these approaches by generating missing components effectively and substantially improving speech quality.

Recently, diffusion-based generative models~\cite{lu2022conditional,richter2023speech,lemercier2023storm} have shown remarkable perceptual quality on processed speech, by re-generating speech signals from random distributions.
For instance, UNIVERSE~\cite{serra2022universe} introduced a score-matching approach to handle a variety of distortions, achieving strong performance in restoring speech.
Afterwards, UNIVERSE++~\cite{scheibler2024universepp} modified the condition model of UNIVERSE to provide high-quality speech features, and it improved stability of the restoration process as well as perceptual quality.
The core of these models is the use of condition modules that adapt to various distortions.
This design enhances restoration performance by incorporating estimated speech features alongside the diffusion model, rather than depending solely on it.

However, generative models often produce hallucinations, especially when provided with inadequate features during sampling or when handling extremely difficult tasks.
Speech restoration is a highly complex task that involves both suppression and generation processes.
When the condition model cannot supply enough informative features, the generative models show limited performance.
In such cases, obtaining condition information from pre-trained models may help reduce performance degradation.
Several works~\cite{hung2022boosting,yue2022reference,byun2023empirical} have confirmed the usefulness of this approach, particularly by leveraging self-supervised learning (SSL)~\cite{baevski2020wav2vec,hsu2021hubert,chen2022wavlm} models or speaker recognition models~\cite{desplanques2020ecapa,jung2022pushing,yakovlev2024reshape} that supply linguistic or speaker-related information for speech restoration.
These condition models, trained through contrastive learning~\cite{oord2018representation,chung2020defence} on real-world speech recordings, can provide content-related information even for distorted speech.

In this paper, we propose a novel approach to conditional speech restoration that leverages auxiliary acoustic cues.
Instead of utilizing content aligned with the clean speech signal being estimated, we employ the acoustic context information present in the input, including environmental attributes.
In particular, we use the CLAP model~\cite{elizalde2023msclap,wu2023laionclap,elizalde2024msclap} to extract embeddings that reflect the environmental attributes in the input.
Since CLAP is trained through correspondence between audio and text captions, it can indicate the soundscape within the signal.
Traditional restoration models implicitly analyze distortions, which sometimes lead to over-suppression or leftover noise depending on the input.
However, by explicitly providing the restoration model with distortion information, our approach minimizes over-suppression and leftover noise, leading to more stable and high-quality restoration results.

Furthermore, we propose an advanced \textit{acoustic context (ACX) representation} that refines CLAP representations.
Although CLAP is effective at describing the soundscape of inputs, it does not fully consider the intensity of each component in input speech.
To distinguish the intensity for more precise acoustic context information, CLAP outputs are embedded into a more detailed latent space.
Simple projection layers are placed onto the CLAP audio encoder while keeping its parameters frozen.
For training, we prepare four different speech samples to learn how to discriminate distortion types and their intensities.
We introduce this novel representation into the restoration model along with CLAP and other content-based conditions.
The following sections provide the details of our context-aware speech restoration and show its effectiveness in speech restoration.
We will demonstrate that this ACX representation is as competitive as conventional representations in restoration tasks, and it has an advantage in more severe environments.

\begin{figure*}[t]
\centering\footnotesize
\hfill
    \begin{minipage}[t]{0.21\linewidth}
        \includegraphics[width=\columnwidth]{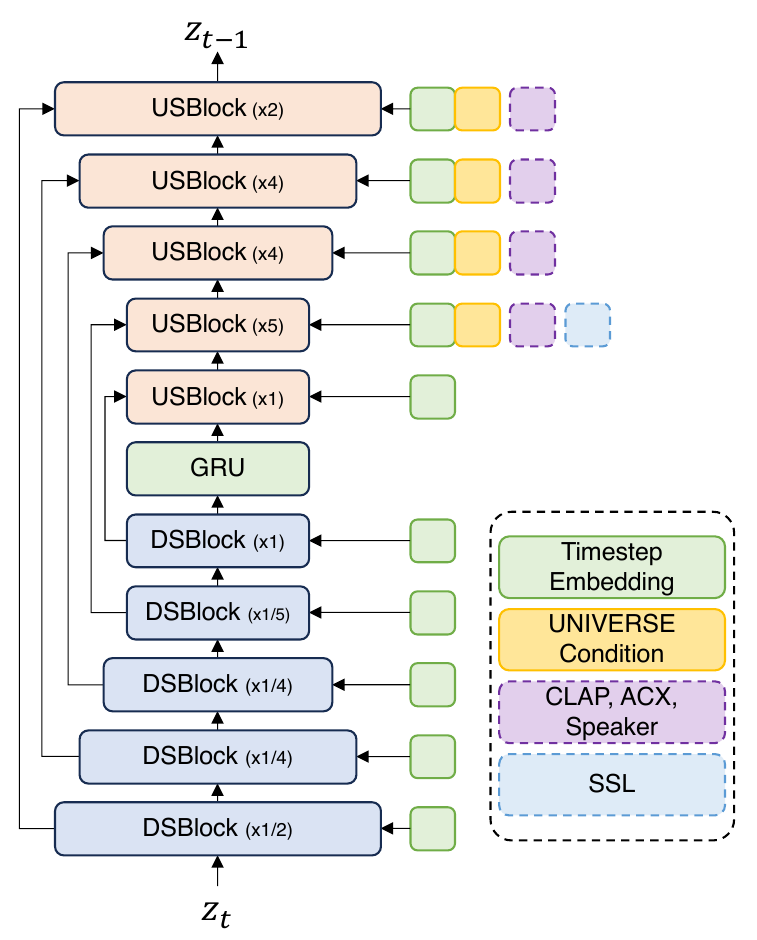}
        \centerline{(a)}
    \end{minipage}
\hfill{\color{gray}\vline}\hfill
    \begin{minipage}[t]{0.48\linewidth}
        \includegraphics[width=\columnwidth]{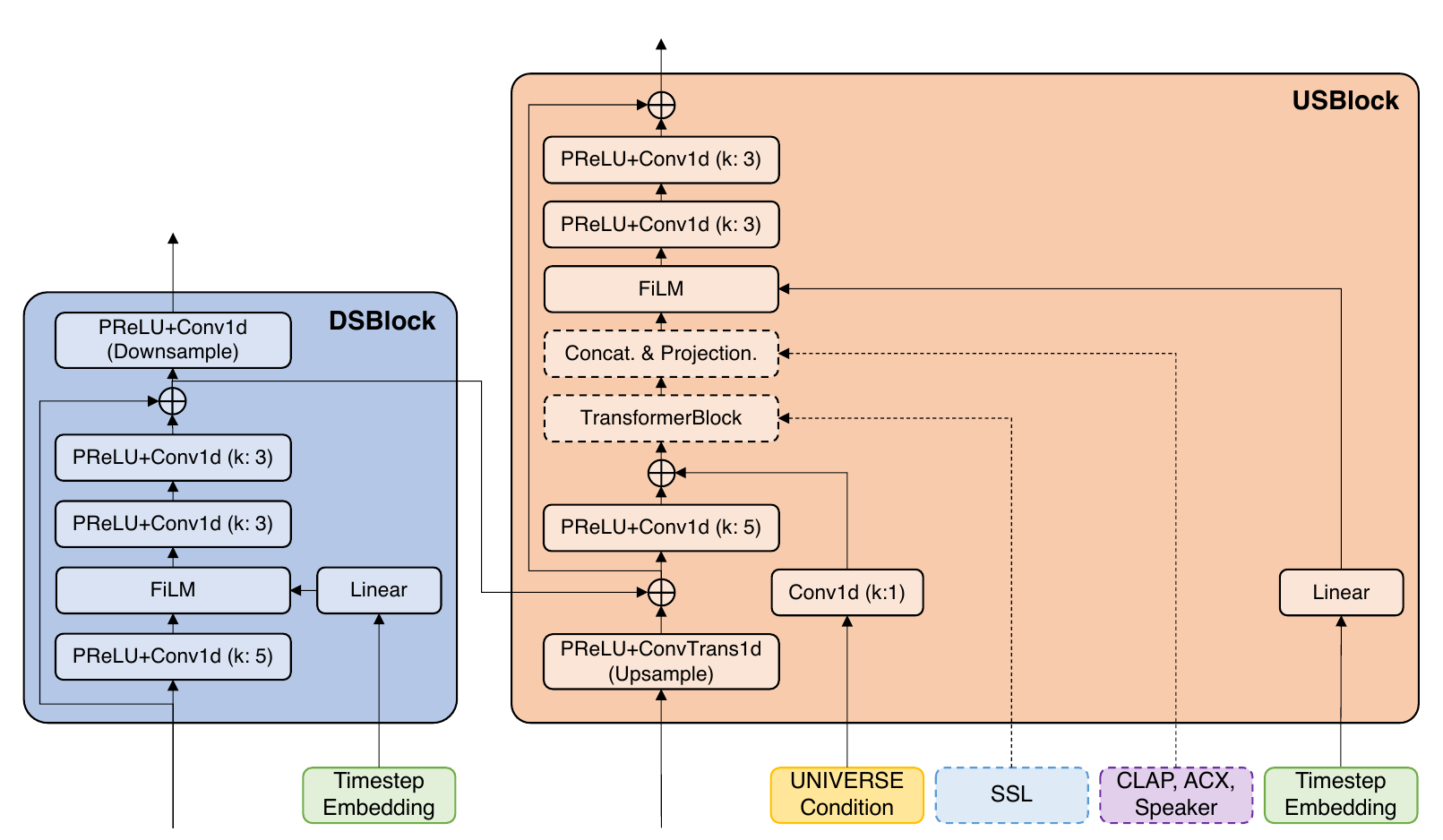}
        \centerline{(b)}
    \end{minipage}
\hfill{\color{gray}\vline}\hfill
    \begin{minipage}[t]{0.10\linewidth}
        \includegraphics[width=\columnwidth]{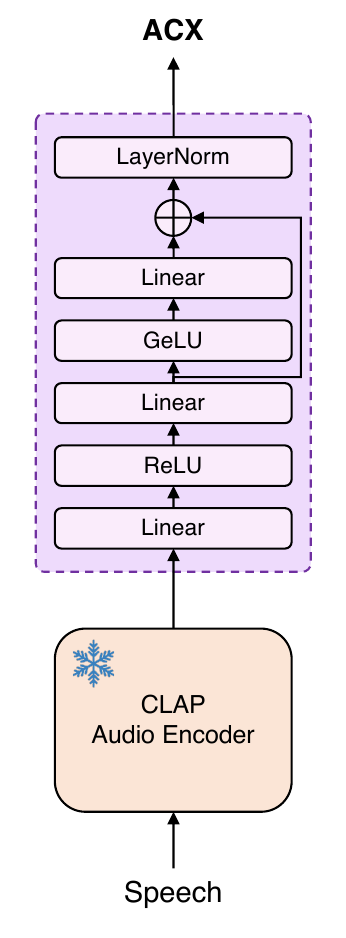}
        \centerline{(c)}
    \end{minipage}
\hfill
\vspace{-7pt}
\xcaption{Illustration of the model architecture. To compare the impact across different conditioning methods, only one condition is enabled at a time (SSL, Speaker, CLAP, ACX). {\normalfont (a)} UNIVERSE++-based diffusion model; {\normalfont (b)} Downsampling (DS) and Upsampling (US) blocks; {\normalfont (c)} ACX model.}
\vspace{-3pt}
\label{fig:block}
\end{figure*}
\section{Related Works}

\xubsection{UNIVERSE \& UNIVERSE++}

UNIVERSE~\cite{serra2022universe} is one of the pioneering methods using a score-based diffusion model for generative speech restoration.
It consists of two networks: the score network and the condition network.
The condition network is trained on multiple mixture density objectives to predict various target features (\eg pitch and harmonicity).
The UNIVERSE model is trained on a wide range of distortions using large-scale datasets, and
it outperforms baseline models in perceptual quality, even with relatively few sampling steps.

UNIVERSE++~\cite{scheibler2024universepp} is an advanced version of UNIVERSE, which modifies the model structure to ensure training stability and introduces adversarial training for the condition network.
Specifically, the condition network adopts a GAN-based strategy to predict clean speech signals instead of relying on the mixture density networks.
The discriminator follows the same recipe as HiFi-GAN~\cite{kong2020hifigan}, incorporating both multi-resolution and multi-period discriminators.
This approach makes the intermediate embeddings of the decoder layers represent high-quality speech features, thereby promoting the score network to predict the target speech signal more efficiently.
As a result, UNIVERSE++ has shown more stable and effective results compared to UNIVERSE and other diffusion-based speech enhancement methods.

\vspace{-3pt}
\xubsection{Contrastive Language-Audio Pretraining (CLAP)}

CLAP~\cite{elizalde2023msclap,wu2023laionclap,elizalde2024msclap} has been in the spotlight for its ability to represent the context of input audio by learning the correspondence between audio and its textual caption.
CLAP is trained via contrastive learning with two encoders, each for audio and text, embedding their respective modalities into a common latent space.
The distance between audio and text embeddings is minimized when they are paired, and maximized otherwise.
For training, large-scale audio datasets~\cite{wu2023laionclap,gemmeke2017audioset} were employed, including various sound sources such as speech, music, ambient sounds, and sound effects.

In audio generation~\cite{liu2023audioldm,majumder2024tango,kong2024audioflamingo}, CLAP has been utilized as a key module to determine the types of sound and their attributes for generated audio.
AudioLDM~\cite{liu2023audioldm,liu2024audioldm2} has been trained using audio embeddings and then generated audio with text prompts from CLAP text encoder, circumventing data scarcity due to the limited availability of paired audio-text datasets.
This demonstrates that CLAP audio embeddings effectively capture the overall acoustic context of input audio, making them a key module for providing environmental context in various generation tasks~\cite{lee2024voiceldm,kim2024astldm,jung2024voicedit}.

\begin{figure*}[t]
    \centering
    \begin{minipage}[t]{\linewidth}
        \centering
        \begin{tikzpicture}
          \begin{axis}[
            hide axis,
            height=0.1\columnwidth,
            xmin=0, xmax=1,
            ymin=0, ymax=1,
            mark options={mark size=1pt, line width=0.5pt},
            legend style={
                nodes={scale=0.6},
                at={(0.5,0.5)},
                anchor=center,
                legend columns=4
                }
          ]
            \addplot[color=Emerald, mark=square] coordinates {(0,0)};
            \addplot[color=Emerald, mark=o] coordinates {(0,0)};
            \addplot[color=RubineRed, mark=square] coordinates {(0,0)};
            \addplot[color=RubineRed, mark=o] coordinates {(0,0)};
            \legend{CLAP:$\mathcal{P}$,CLAP:$\mathcal{N}$,\shortname:$\mathcal{P}$,\shortname:$\mathcal{N}$}
            \end{axis}
        \end{tikzpicture}
    \end{minipage}
    \vfill
    \hfill
    \begin{minipage}[t]{0.31\linewidth}
        \centering\scriptsize
        \begin{tikzpicture}
            \begin{axis}[
                width=\columnwidth, 
                height=.7\columnwidth,
                scaled y ticks = false,
                scaled x ticks = false,
                xtick pos=left, ytick pos=left,
                bar width=5pt,
                enlargelimits=0.005,
                xmin=-1, xmax=31,
                ymin=-0.5, ymax=1.0,
                ylabel={Cosine Similarity},
                ylabel near ticks,
                xlabel={(a) SNR (dB)},
                xlabel near ticks,
                mark options={mark size=1pt, line width=0.5pt},
                ytick={-0.4, -0.2, 0.0, 0.2, 0.4, 0.6, 0.8, 1.0},
                yticklabels={-0.4, -0.2, 0.0, 0.2, 0.4, 0.6, 0.8, 1.0},
                minor tick length=1ex,
                x tick label style={/pgf/number format/1000 sep=},
                ymajorgrids=true,
                xmajorgrids=true,
                grid style=dashed,
                legend style={nodes={scale=0.6}, anchor=south, at={(0.5, -0.5)}, legend columns=4},
                ]
                \addplot[color=Emerald, mark=square] coordinates{
(0,0.846778591208666) (2,0.842824165031169) (4,0.840156128901301) (6,0.838243242968055) (8,0.836966937028089) (10,0.836358764290231) (12,0.836151220338437) (14,0.836106360682006) (16,0.836072097125563) (18,0.836203323188916) (20,0.836910347553712) (22,0.838330177716839) (24,0.840602827737632) (26,0.843829108384049) (28,0.847863094031232) (30,0.852046470546607)};
                \addplot[color=Emerald, mark=o] coordinates{
                (0,0.401922796608754) (2,0.433879322851601) (4,0.466192012725091) (6,0.498118964213769) (8,0.528769040353669) (10,0.557559409585682) (12,0.584319155156902) (14,0.609233098737534) (16,0.632848529850395) (18,0.656012343731031) (20,0.679642647794149) (22,0.703951770096149) (24,0.728187711855162) (26,0.751487285983794) (28,0.773543503099275) (30,0.793761098920142)};
                \addplot[color=RubineRed, mark=square] coordinates{
                (0,0.831615811702117) (2,0.83349242987274) (4,0.838007817786295) (6,0.843950175255248) (8,0.850176886158082) (10,0.855342907885325) (12,0.858409846291959) (14,0.85862903644159) (16,0.855726317909447) (18,0.849707229551181) (20,0.841351232627063) (22,0.832161827103316) (24,0.824351830733488) (26,0.819327834812762) (28,0.819521596130816) (30,0.829668296410621)};
                \addplot[color=RubineRed, mark=o] coordinates{
                (0,-0.26052478387007) (2,-0.254838755909236) (4,-0.261397138115887) (6,-0.2792386411965) (8,-0.30237663257415) (10,-0.323262054978382) (12,-0.33271022270732) (14,-0.320025101833009) (16,-0.27910575129146) (18,-0.209365630694168) (20,-0.114663070155923) (22,0.00221356145531228) (24,0.138458021500519) (26,0.291368031101211) (28,0.452848269769162) (30,0.606897811852938)};
            \end{axis}
        \end{tikzpicture}
    \end{minipage}%
    \hfill
    \begin{minipage}[t]{0.31\linewidth}
        \centering\scriptsize
        \begin{tikzpicture}
            \begin{axis}[
                width=\columnwidth,  
                height=.7\columnwidth,
                scaled y ticks = false,
                scaled x ticks = false,
                xtick pos=left, ytick pos=left,
                bar width=5pt,
                enlargelimits=0.005,
                xmin=4, xmax=10,
                ymin=0.1, ymax=1.0,
                ylabel={Cosine Similarity},
                ylabel near ticks,
                xlabel={(b) Room size},
                xticklabels={Large, Medium, Small},
                xlabel near ticks,
                mark options={mark size=1pt, line width=0.5pt},
                xtick={5,7,9},
                ytick={0.0, 0.2, 0.4, 0.6, 0.8, 1.0},
                yticklabels={0.0, 0.2, 0.4, 0.6, 0.8, 1.0},
                minor tick length=1ex,
                x tick label style={/pgf/number format/1000 sep=},
                ymajorgrids=true,
                xmajorgrids=true,
                grid style=dashed,
                legend style={nodes={scale=0.6}, anchor=south, at={(0.5, -0.5)}, legend columns=4},
                ]
                \addplot[color=Emerald, mark=square] coordinates{ 
                (5, 0.8616972613392524) (7, 0.8545544883869227) (9, 0.8606791108557321)};
                \addplot[color=Emerald, mark=o] coordinates{ 
                (5, 0.8123121527792181) (7, 0.7976171270880884) (9, 0.8136362902892446)};
                \addplot[color=RubineRed, mark=square] coordinates{ 
                (5, 0.8744041338852309) (7, 0.8727271190112077) (9, 0.9030545117351615)};
                \addplot[color=RubineRed, mark=o] coordinates{ 
                (5, 0.24561177364785622) (7, 0.34261168382419904) (9, 0.6716271367972916)};
            \end{axis}
        \end{tikzpicture}
    \end{minipage}%
    \hfill
    \begin{minipage}[t]{0.31\linewidth}
        \centering\scriptsize
        \begin{tikzpicture}
            \begin{axis}[
                width=\columnwidth,  
                height=.7\columnwidth,
                scaled y ticks = false,
                scaled x ticks = false,
                xtick pos=left, ytick pos=left,
                bar width=3pt,
                enlargelimits=0.005,
                xmin=0, xmax=9,
                ymin=0.5, ymax=1.0,
                ylabel={Cosine Similarity},
                ylabel near ticks,
                xlabel={(c) Cutoff frequency (kHz)},
                xlabel near ticks,
                mark options={mark size=1pt, line width=0.5pt},
                ytick={0.0, 0.2, 0.4, 0.6, 0.8, 1.0},
                yticklabels={0.0, 0.2, 0.4, 0.6, 0.8, 1.0},
                minor tick length=1ex,
                x tick label style={/pgf/number format/1000 sep=},
                ymajorgrids=true,
                xmajorgrids=true,
                grid style=dashed,
                legend style={nodes={scale=0.6}, anchor=south, at={(0.5, -0.5)}, legend columns=4},
                ]
                \addplot[color=Emerald, mark=square] coordinates{ 
                (1,0.839715669456038) (2,0.900200233731455) (3,0.886778339888286) (4,0.895296706664331) (5,0.899040883750591) (6,0.893658303520055) (7,0.887469804258023) (8,0.871977504234291)};
                \addplot[color=Emerald, mark=o] coordinates{ 
                (1,0.709574533825361) (2,0.767472237999578) (3,0.805239231551735) (4,0.846389164913048) (5,0.859384010211357) (6,0.870988318642366) (7,0.873150004489908) (8,0.871977504234291)};
                \addplot[color=RubineRed, mark=square] coordinates{ 
                (1,0.949879240497802) (2,0.979655615913058) (3,0.977676686008) (4,0.982502397330641) (5,0.983015647935636) (6,0.979981585686068) (7,0.976367517076071) (8,0.97519703433641)};
                \addplot[color=RubineRed, mark=o] coordinates{ 
                (1,0.574505054672222) (2,0.675796799470209) (3,0.786079352703488) (4,0.816687254651079) (5,0.856614827386384) (6,0.913435448865289) (7,0.939217224427797) (8,0.97519703433641)};
            \end{axis}
        \end{tikzpicture}
    \end{minipage}%
    \hfill
    \vspace{-11pt}
    \xcaption{Cosine similarity of the context representations between anchors and positive samples $\mathcal{P}$ (with the same distortion type and intensity) and between anchors and negative samples $\mathcal{N}$ (clean speech) for three types of distortion: {\normalfont (a)} Noise, {\normalfont (b)} Reverberation, and {\normalfont (c)} Bandlimiting distortion.}
    \vspace{-4pt}
    \label{fig:acxr}
\end{figure*}

\section{Proposed Method}

\xubsection{Context-aware Speech Restoration}
We propose a novel approach for conditional restoration that leverages acoustic context information.
Unlike content-aware methods that mainly rely on speaker information or linguistic cues~\cite{hung2022boosting,yue2022reference}, our approach focuses on the acoustic context in the input speech.
Since the CLAP audio encoder is trained on audio captions describing various acoustic scenes,
it can effectively capture different sound sources, which makes it well-suited for providing acoustic context in the restoration model.
Therefore, we use this audio encoder to supply acoustic context information for the restoration instead of content-related information.
We specifically integrate this approach into UNIVERSE++, which has shown impressive restoration performance.
Although UNIVERSE++ excels at speech restoration, it still struggles under harsh conditions that degrade its conditional features.
The acoustic context information would improve overall restoration performance, especially under severe distortions.
Hence, we slightly modify the structure of UNIVERSE++ to integrate extra conditional features, as shown in \Fig{block}(a) and (b), and investigate the impact of this conditioning strategy.

\vspace{-2pt}
\xubsection{Acoustic Context (ACX) Representation}
We also propose an advanced acoustic context representation built on CLAP embeddings.
Although CLAP effectively represents the overall acoustic context, it does not fully capture the severity of environmental distortions present in speech signals.
For instance, while it can identify ambient sounds or distinguish whether the speaker is indoors or outdoors, it fails to quantify noise or reverberation levels.
Furthermore, because CLAP embeddings capture every sound in the input, they also include speaker-related attributes such as gender or age, which are irrelevant to the distortions we aim to remove.

To address these issues, we incorporate metric learning strategy to obtain more precise ACX representation.
Our strategy aims to learn discriminability not only for different acoustic context types, but also for their intensity, thereby capturing fine-grained variations in distorted inputs.
We add extra layers atop the pretrained CLAP encoder to project its embeddings onto another latent space as shown in \Fig{block}(c).
For training, we prepare four different audio samples: anchor ($\mathcal{A}$), positive ($\mathcal{P}$), weak negative ($\mathcal{N}_w$), and hard negative ($\mathcal{N}_h$).
The anchor and the positive audio share identical acoustic conditions including noise type, signal-to-noise ratio (SNR), room impulse response (RIR), and cutoff frequency, yet they differ in speech content.
In contrast, the weak negative consists of different distortions compared to the anchor, thus encouraging the representation to learn how to discriminate among different acoustic contexts.
For training efficiency, we exploit other samples within a minibatch as the weak negatives.
The key sample in training is the hard negative, which pushes the representation to provide a fine-grained description according to the intensity of the distortions.
Hard negatives are formed by reusing anchor or positive speech with the same distortion type but at a different intensity level from the positive sample.
With this strategy, we reduce the attributes about spoken terms on the embeddings.

We design the training criteria using both the L2 distance $d$ and the cosine similarity $s$ between embeddings. 
For discriminability, the anchor–positive distance is minimized and the weak negative is pushed away from the anchor.
Likewise, the anchor–positive cosine similarity is maximized, while the anchor–weak negative similarity is minimized as follows:\vspace{-2pt}
\begin{equation}\scriptsize
\label{eq:discriminability}
    \begin{gathered}
        L_c=-\sum_{n\in{\mathcal{A}}}\sum_{m\in{\{\mathcal{P},\mathcal{N}_w}\}}y_{m}\text{log}\biggl(\frac{\exp(s_{n,m})}{\sum_{m\in{\{\mathcal{P},\mathcal{N}_w}\}}\exp(s_{n,m})}\biggr)\\
        L_d=-\sum_{n\in{\mathcal{A}}}\sum_{m\in{\{\mathcal{P},\mathcal{N}_w}\}}y_{m}\text{log}\biggl(\frac{\exp(d_{n,m}^{-1})}{\sum_{m\in{\{\mathcal{P},\mathcal{N}_w}\}}\exp(d_{n,m}^{-1})}\biggr)\\
        y_{m}=
        \begin{cases}
            1,& m\in\mathcal{P}\\
            0,& m\in\mathcal{N}_w
        \end{cases},
    \end{gathered}
\end{equation}
where $s_{x,y}=\cos(\phi(x), \phi(y))$ and $d_{x,y}=||\phi(x)-\phi(y)||_2$. $\phi(\cdot)$ is the ACX representation model in \Fig{block}(c).

To learn the intensity, the hard negative is treated differently from the weak negative.
The hard negative embedding should exhibit higher similarity with the anchor than the weak negative.
Nevertheless, it remains farther from the anchor than the positive, thus preventing it from becoming indistinguishable from the positive sample.
Therefore, we maximize the similarity between the anchor and the hard negative embedding, but cap it at the maximum similarity observed for the positive embedding.
Also, we do not minimize the anchor–hard negative distance but only maximize distance among negatives.
The training criteria for the intensity are given as follows:\vspace{-2pt}
\begin{equation}\scriptsize
\label{eq:intensity}
    \begin{gathered}
        L_{nd}=-\sum_{n\in{\mathcal{N}_w}}\sum_{m\in{\mathcal{N}_h}}d_{n,m} ,\\
        L_{nc}=-\sum_{n\in{\mathcal{A}}}\sum_{m\in{\{\mathcal{N}_{h},\mathcal{N}_w}\}}y_{m}\,\text{log}\Bigl(\frac{\exp(\hat{s}_{n,m})}{\sum_{m\in{\{\mathcal{N}_{h},\mathcal{N}_w}\}}\exp(\hat{s}_{n,m})}\Bigr),\\
        y_{m}=
        \begin{cases}
            1,& m\in\mathcal{N}_h\\
            0,& m\in\mathcal{N}_w
        \end{cases},
        \hat{s}_{n,m}=
         \begin{cases}
            \min(s_{n,m}, p_{max}),& m\in\mathcal{N}_h\\
            s_{n,m},& m\in\mathcal{N}_w
        \end{cases},
    \end{gathered}
\end{equation}
where $p_{max}$ is the maximum similarity score between the anchor and the positive within the minibatch.
We trained the \shortname~model by integrating training criteria as follows,
\begin{equation}\footnotesize
    L=L_c+L_d+L_{nd}+L_{nc}.
\end{equation}
During training, the CLAP encoder is frozen to preserve its contextual capacity.

\begin{table*}[t]
    \centering
    \scriptsize
    \begin{tabular}{ l | c | c | c | c | c | c | c | c | c | c | c | c | c | c }
        \toprule
        \multirow{2}{*}{\bf Model} & \multicolumn{6}{c|}{\bf SNR (dB)}  & \multicolumn{4}{c|}{\bf Cutoff Frequency (kHz)} & \multicolumn{3}{c|}{\bf Room Size} & \multirow{2}{*}{\bf Average} \\ 
        \cmidrule{2-14}
            & -5        & 0         & 5         & 10        & 15        & 20        & 1k        &3k         &5k         &7k         & Large     & Medium    & Small     & \\
         \midrule
UNIVERSE    & 1.823     & 1.991     & 2.105     & 2.155     & 2.175     & 2.186     & 1.931     & 2.079     & 2.128     & 2.143     & 2.219     & 2.227     & 2.254     & 2.109 \\ \midrule
UNIVERSE++  & 2.118     & 2.358     & 2.515     & 2.594     & 2.646     & 2.659     & 2.232     & 2.529     & 2.553     & 2.597     & 2.787     & 2.836     & 2.862     & 2.560 \\
~~~+HuBERT  & 2.117     & 2.392     & 2.565     & 2.641     & 2.685     & 2.701     & \bf 2.301 & 2.547     & 2.620     & 2.633     & \bf 2.878 & \bf 2.921 & \bf 2.951 & 2.612 \\
~~~+RawNet3 & 2.166     & 2.395     & 2.561     & 2.631     & 2.684     & 2.704     & 2.269     & 2.584     & 2.630     & 2.640     & 2.849     & 2.891     & 2.918     & 2.609 \\
~~~\bf +CLAP    & \bf 2.168     & \bf 2.426 & 2.566     & 2.648 & \bf 2.712 & \bf 2.721 & 2.244     & \bf 2.610 & 2.641     & 2.646     & 2.847     & 2.886     & 2.922     & 2.618 \\
~~~\bf +ACX & 2.167 & 2.411 &\bf 2.576 & \bf 2.649 & 2.702 & 2.718 & 2.230 & 2.598 & \bf 2.642 & \bf 2.666 & 2.865 & 2.908 & 2.930 & \bf 2.620 \\

\bottomrule
    \end{tabular}
    \tcaption{DNSMOS results across various evaluation sets at different distortion intensities. Each dataset in a column has a fixed intensity for the corresponding distortion, while other distortions remain random.}
    \vspace{-10pt}
\label{tab:intensity}
\end{table*}

\vspace{-5pt}
\section{Experiments}
\xubsection{Implementation Details}\vspace{-2pt}

\paragraph{Datasets}
We consider four environmental factors for experiments: noise, reverberation, bandlimiting, and clipping.
We use VCTK~\cite{yamagishi2019vctk} and LibriSpeech~\cite{panayotov2015librispeech} as speech corpora, DEMAND~\cite{thiemann2013demand} and WHAM~\cite{wichern2019wham} for noise, and DNS-Challenge~\cite{dubey2023dnschallenge} and DiffRIR~\cite{wang2024diffrir} for room acoustics.
For datasets with predefined training and test splits, we follow their standard configurations.
For VCTK and DEMAND, the training and test samples are selected based on the Valentini dataset~\cite{valentini2017noisy}.
To evaluate reverberation, we randomly select 10 RIR samples per room type from DNS-Challenge and 5 per room type from DiffRIR for the test set.
Bandlimiting distortions are simulated using resampling and low-pass filtering, applying a random cutoff frequency between 1kHz and 8kHz.
For clipping, we randomly clamp speech amplitudes by up to 50\% to simulate distortion.
For structured evaluation, 13 test subsets are created, each with 1,000 speech samples, isolating a specific distortion level while applying others randomly.
SNR levels range from -5dB to 20dB in 5dB steps, reverberation is categorized by room size, and cutoff frequency is set from 1kHz to 7kHz in 2kHz intervals.
Additionally, restoration performances are evaluated using the URGENT challenge dataset~\cite{zhang2024urgent}, which includes a broader range of distortions.

\paragraph{Evaluation metrics}
We adopt non-intrusive metrics to assess the perceptual quality of restored speech.
Specifically, DNSMOS~\cite{reddy2022dnsmos} and SIGMOS~\cite{ristea2024sigmos}, both neural quality prediction models, are used for quantitative evaluation.
Also, to check the characteristics of the ACX representation, we measure cosine similarity between embeddings.

\paragraph{Conditions}
We evaluate four types of conditioning: speaker embeddings, SSL features, CLAP embeddings, and ACX representation.
RawNet3~\cite{jung2022pushing} for speaker embeddings is used, which outputs 256-dimensional vectors.
Among various SSL models, we choose HuBERT-base~\cite{hsu2021hubert} and extract 768-dimensional framewise embeddings from its 9th layer.
Both CLAP\scriptnote{\url{https://huggingface.co/microsoft/msclap}}~\cite{elizalde2024msclap} and our proposed \shortname\ representation produce a 1,024-dimensional embedding per utterance.

\paragraph{Model structure}
Experiments are built on UNIVERSE++\scriptnote{\url{https://github.com/line/open-universe}}, modifying input length to 5.12 seconds.
\Fig{block}(a) and (b) show a detailed illustration of the diffusion network.
Global conditions are integrated through simple concatenation to avoid increasing the model size.
Specifically, we integrate utterance-wise conditions (\ie Speaker, CLAP, \shortname) into the input of each decoder layer of UNIVERSE++ by concatenation followed by a linear projection.
SSL features are fed into the second decoder layer via a transformer decoder structure.
On top of the CLAP audio encoder, we stack two fully-connected layers and a projection layer to extract ACX representation, as shown in \Fig{block}(c).

\xubsection{Experiment Results}\vspace{-2pt}

\paragraph{Representation Performances}
We first evaluated the \shortname\ representation to assess its ability to differentiate various distortions and capture their intensity.
To do this, we prepared three test sets: noise, reverberation, and bandwidth limitation.
Each set contains 823 samples with the same distortion type and intensity.
For noise, SNR levels were tested from 0 to 30dB at 1dB intervals.
For reverberation, we used RIR samples categorized as small, medium, and large rooms in the DNS-Challenge dataset.
For bandlimit distortion, we progressively reduced the bandwidth from 8kHz down to 1kHz.

\Fig{acxr} shows the cosine similarity between the anchor and clean sample ($\mathcal{N}$) and between the anchor and positive sample ($\mathcal{P}$).
If the similarity for $\mathcal{N}$ decreases as distortion worsens while the similarity for 
$\mathcal{P}$ remains high, the embedding can be considered effective in representing both distortion type and intensity.
Otherwise, if both similarities stay high, the embedding fails to capture intensity variations.
In \Fig{acxr}, we observed that CLAP embeddings already exhibit moderate discriminability and intensity-awareness.
However, ACX shows a larger drop in similarity as distortions become more severe, indicating that it represents distortion intensity more precisely than CLAP across all distortion types.

\begin{table}[t]
    \centering
    \scriptsize
    \begin{tabular}{ l | c | c | c | c}
        \toprule
        \multirow{2}{*}{\bf Model}  & \multicolumn{2}{c|}{\bf Blind} & \multicolumn{2}{c}{\bf NonBlind} \\ 
          &   
         \bf DNSMOS & \bf SIGMOS & \bf DNSMOS & \bf SIGMOS \\
\midrule
UNIVERSE    & 2.088     & 2.432     & 2.356     & 2.563 \\
\midrule
UNIVERSE++  & 2.506     & 2.463     & 3.002     & 2.858 \\
~~~+HuBERT  & 2.579     & 2.567     & 3.040     & \bf 2.944 \\
~~~+RawNet3 & 2.560     & 2.561     & 3.028     & 2.904 \\
~~~\bf +CLAP    & 2.570     & 2.519     & 3.019     & 2.849 \\
~~~\bf +ACX     & \bf 2.583     & \bf 2.583 & \bf 3.053     & 2.940 \\
         \bottomrule
    \end{tabular}
    \tcaption{Evaluation results on URGENT-Challenge datasets.}
    \vspace{-12pt}
    \label{tab:urgent}
\end{table} 
\paragraph{Restoration Performances}
\Tab{intensity} presents DNSMOS results across test sets with varying distortion intensities.
Compared to UNIVERSE and UNIVERSE++, models with additional conditioning achieve better overall performance.
CLAP and ACX outperform content-aware restoration methods in most cases, except in the reverberation subset, where HuBERT-based models achieve higher scores.
This is primarily because reverberation has minimal impact on linguistic content, allowing HuBERT to provide effective guidance.
However, in other scenarios, models utilizing context information demonstrate clear advantages.
Both CLAP and ACX outperform RawNet3, which relies on speaker-based conditioning.
Unlike content-aware approaches, which remove unmatched components from input speech, context-aware methods explicitly incorporate distortion information to support the restoration task.
When comparing CLAP and ACX, there is little difference in overall quality, but ACX outperforms CLAP in reverberation subsets.
This is because CLAP struggles to distinguish room acoustics, whereas ACX, as shown in \Fig{acxr}, effectively captures the intensity of reverberation, providing better conditioning for the model.

\Tab{urgent} reports the restoration performance on the URGENT challenge dataset.
Although these samples include distortion types unseen during training, context-based representations demonstrate strong performance, even in comparison to content-based approaches.
The speech samples in this dataset contain moderate distortions that minimally obscure information relevant to speech content.
As a result, content-based models, especially those using HuBERT, achieve strong performance.
Nonetheless, ACX performs best on the blind set, where restoration is more challenging.
This indicates that ACX provides more reliable guidance across diverse real-world scenarios compared to other conditioning approaches.


\begin{table}[t]
    \centering
    \scriptsize
    \begin{tabular}{ l | c | c | c | c}
        \toprule
        \multirow{2}{*}{\bf Model}  & \multicolumn{2}{c|}{\bf DNSMOS} & \multicolumn{2}{c}{\bf SIGMOS} \\ 
          &   
         \bf Mean & \bf Std. & \bf Mean & \bf Std. \\
         \midrule
         UNIVERSE           & 2.109       & \bf 0.373   & 2.424         & \bf 0.366 \\
         \midrule
         UNIVERSE++         & 2.560       & \bf 0.520   & 2.311         & 0.549 \\
        ~~~+HuBERT          & 2.612       & 0.570       & 2.408         & 0.574 \\
        ~~~+RawNet3         & 2.609       & 0.544       & 2.461         & 0.574 \\
        ~~~\bf +CLAP            & 2.618       & 0.521       & 2.409         & 0.548 \\
        ~~~\bf +ACX             & \bf 2.620   & 0.531       & \bf 2.491     & \bf 0.544 \\
         \bottomrule
    \end{tabular}
    \tcaption{Mean and standard deviation of DNSMOS and SIGMOS on restoration results.}
    \vspace{-10pt}
    \label{tab:ablation}
\end{table}

\paragraph{Stability of Restoration Performances}
While restoration quality naturally degrades as distortions worsen, maintaining stable performance across all inputs remains essential.
To assess this stability, performance variations were analyzed by reporting the mean and standard deviation of DNSMOS and SIGMOS for each conditioning setup, as shown in~\Tab{ablation}.
UNIVERSE and UNIVERSE++ show relatively low variance due to the absence of additional conditioning models, but their overall performance remains limited.
HuBERT achieves high restoration quality but exhibits large variability, as its framewise feature extraction makes it sensitive to local distortions, leading to unstable conditioning results.
In contrast, RawNet, CLAP, and ACX offer stable utterance-level features, delivering consistent performance across datasets.
ACX, in particular, adapts to different environments by leveraging distortion level information, further minimizing performance variations across datasets.
These results confirm that ACX is a highly effective conditioning method, offering both strong performance and stability.

\vspace{-5pt}\section{Conclusion}\vspace{-3pt}

In this study, we introduced acoustic context as a novel conditioning strategy in speech restoration task.
We first utilized the CLAP model to capture environmental attributes and further refined its embeddings into acoustic context (ACX) representation, allowing the model to represent both distortion type and intensity.
We applied these proposed conditions to UNIVERSE++ model and compared them with other content-related conditions, HuBERT and RawNet3, which respectively supply linguistic and speaker-related information.
Experimental results demonstrated that acoustic context information achieves competitive or superior performance in restoration tasks while improving stability.
These findings highlight the effectiveness of ACX, and future work will explore integrating context and content-based features for further improvements.


\bibliographystyle{IEEEtran}
\bibliography{shortstrings,mybib}

\end{document}